\documentstyle[twocolumn,aps,epsf]{revtex}
\def\mb#1{\mbox{\boldmath $#1$}}

\begin{document}

\draft
\preprint{HEP/123-qed}
\title{Random Lasing from Weakly Scattering Media: Universality in
the Emission Spectra from  $\pi$-Conjugated Polymer Films}
\author{R. C. Polson, M. E. Raikh, and Z. V. Vardeny }
\address{Department of Physics, University of Utah, Salt Lake City, 
Utah 84112}

\date{\today}
\maketitle

\begin{abstract}
When films of  $\pi$-conjugated polymers are optically excited above a
certain threshold intensity,
then the emission spectrum acquires a multimode 
finely structured shape, which 
depends on the position of the excitation spot. We demonstrate that
the power Fourier transform  (PFT) of the emission spectrum exhibits a
certain peak-like structure, which  also depends on the excitation 
spot. Our intriguing observation is that 
{\em averaging} the individual PFTs does
not lead  to a {\em structureless} curve, but rather yields a series of
 {\em distinct} transform peaks.  This suggests {\em universality},
 namely that the underlying random resonators that are responsible
 for the laser emission from the $\pi$-conjugated polymer film 
are almost
 {\em identical}.  We argue that the reason for such an universality
 is the large size of a typical resonator, which we determined from
 the PFT, as compared to the emission wavelength,
 $\lambda$.  This fact is, in turn, a consequence of the large light
 mean free path, $l^{\ast}\simeq 10 \lambda$ in the polymer film.  This
 contrasts previous observations of random lasing in powders, where 
 $l^{\ast}\sim \lambda$.  
We develop a simple theory that
 explains the presence of peaks in the average PFT
 and predicts their shape.  The results of the theory agree
 {\em quantitatively} with the data.
\end{abstract}



\section{Introduction} 

Random lasing is a young and rapidly growing area of research.  In
fact, the phenomena ``random lasing'' comprise two distinctively 
different
subfields of thorough theoretical and experimental studies.  
We  dub them here
``incoherent'' random lasing  and ``coherent''random lasing, 
respectively.  The essence of both subfields is the
propagation of light in random media with gain.  

{\em Incoherent random lasing} corresponds to a situation where, 
in the absence of a resonator,
multiple scattering  prevents light from leaving the optically
pumped region. Such a disorder-induced feedback is sufficient for a
drastic narrowing of the emission spectrum
 above a certain threshold pump
intensity. Research on incoherent random lasing addresses various
aspects such as steady state spectral characteristics and dynamics of
this narrowing.  {\em Coherent random lasing}, on the contrary
 pursues a scenario where
disorder in a random medium leads to complete (or nearly complete)
localization of the photonic modes. In other words, the disorder assumes
the role of a Fabry-Perot resonator in a conventional laser.
Therefore, above a threshold excitation intensity
 the emission spectrum from 
such a disordered
medium comprises of  a number of very sharp, laser-like lines.

The message of the present paper is that within the 
subfield of coherent random
lasing there exist two different regimes, which we
dub here ``quantum'' and ``classical''.  
These two regimes are analogous
to the quantum (Anderson) and classical localization of electrons in a
random potential. If the potential is short-range, then the only length
scale associated with it is the elastic mean free path,
$l^{\ast}$. Increasing   disorder decreases the value of
$l^{\ast}$.  Anderson localization is expected when $l^{\ast}$ becomes
smaller than $\lambda$, where $\lambda$ is the electronic wavelength.
In contrast to the short-range potential, the long-range potential 
can be envisioned as a smooth landscape consisting of
``lakes'' and ``mountains''. If a typical
spatial scale of this potential is much bigger than $\lambda$,
then it is
obvious that  electrons with energies near the bottom of each
lake are localized. The origin of this localization is purely
classical.  Interference effects, which are crucial for
 Anderson localization play a secondary role
in the classical localization scenario; they are responsible only for
the {\em exact} positions of
the energy levels inside the lakes. Within this picture, the
localization length of the electronic states is of the order of
the lake diameter, and thus much bigger than $\lambda$.  

Since there is no one-to-one correspondence between the
Schr\"odinger equation  and the wave equation, 
the above picture of the classical localization
cannot be directly applied to  light. Nevertheless, as we 
argue below, the long-range fluctuations of the refraction
index, $n(\mb{\rho})$ are able to trap the light. The analog of a lake
in optics is a region with higher $n(\mb{\rho})$ than the average
$n$ 
of the surrounding space. Classical localization is then due to the
total internal reflections from the boundaries of this
region, as follows from geometrical optics.

In this paper we  present  evidence that coherent random
lasing from $\pi$-conjugated  polymer
films is of the {\em classical type}. The basic argument in favor of
this conclusion is that a long-range
fluctuation of $n(\mb{\rho})$ is similar to
a Fabry-Perot resonator, in the sense that it
gives rise to  a {\em number}
of localized modes having close frequencies and close quality factors.
These modes are revealed  in the emission spectrum.
Since the frequencies of such modes are correlated, we could
estimate the size, $L$ of the underlying resonators from the   
power Fourier transform (PFT) of the  spectrum. We indeed
find that $L\gg\lambda$. 

\begin{figure}
\narrowtext
{\epsfxsize=8.5cm
\centerline{\epsfbox{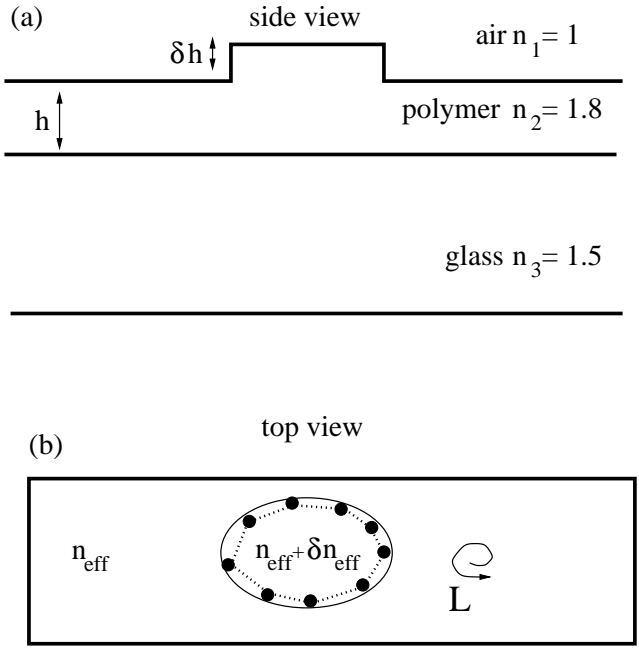}}}
\protect\vspace{0.5cm}
\caption{(a) Schematic illustration of the thickness fluctuation of the polymer film; the refractive indices of the polymer and glass substrate are given. (b) Schematic illustration of the whispering-gallery type mode in a resonator formed by a thickness variation $\delta h$; $L$ is the round trip resonator pathlength.}  
\label{F:cartoon}
\end{figure}
\noindent

The most probable reason for the fluctuations of $n(\mb{\rho})$
in the polymer films is inhomogeneity of the film thickness, $h$.
Since light is confined within the film due to 
waveguiding, (see Fig. 1a) larger $h$ leads to higher 
in-plane wave vector 
of the light modes, {\em i.e.} to  higher {\em effective} refraction
index. By virtue of this mechanism, the long-range fluctuation of
$h$ can result in the formation of a microdisk-type resonator 
(Fig. 1b). Such resonators have recently attracted a lot of 
attention\cite{slusher,gmachl},  since they constitute the 
key element of semiconductor microlasers.

 Certainly, alongside with long-range
$n(\mb{\rho})$ fluctuations, short-range disorder 
is also present in the polymer films. 
This leads to $l^{\ast}\simeq 10 \lambda$,
which we determined from coherent backscattering measurements.
Such a weak short-range disorder  is unable to localize 
light by itself. However, it radically affects  lasing from
the polymer film. This is because, whereas $l^{\ast}\gg \lambda$,
it is still much shorter than the resonator size, {\em i.e.}
$l^{\ast}\ll L$; this
means that the ability of most random resonators 
to trap  light is suppressed
by the short-range disorder. There is a dramatic consequence
of such a suppression: those resonators that ``survive''
the short-range disorder are
sparse, and consequently {\em almost identical}. This scenario
manifests itself in a spectacular way.
At each position of the excitation spot the PFT 
of the random emission spectrum exhibits certain features.
We found that averaging 
the PFTs of individual random spectra 
over the positions of the excitation spot on the polymer film
 not only does not
smear these features, but, on the contrary, yields a 
series of {\em distinct} transform peaks.
Moreover, we found that the shape of the averaged PFT is
{\em universal}, {\em i.e.} increasing the disorder and correspondingly
reducing $l^{\ast}$ does not change  this shape; the average of the PFT
spectra at different $l^{\ast}$ scales with $l^{\ast}$ to
a universal curve.  We have also developed a simple theory for the
the PFT. Without specifying the shape of the
resonators, the theory  is based on a {\em single} assumption that
the random resonators are {\em exponentially} sparse. 
This assumption is sufficient to reproduce the universal shape
of the average  PFT.

The paper is organized as follows. In Sect.~II we review
the current state of the theory and the  experiment in the subfields
of coherent and incoherent random lasing.  
In Sect.~III we report on our measurements of random lasing spectra,
PFTs and spatial intensity distribution of the emission
from $DOO-PPV$ polymer films. In Sect.~IV a simple theory
of random resonators formed due to long-range fluctuations
of $n(\mb{\rho})$ is presented. Sect.~V concludes the paper.


\section{Review of Random Lasing}

\subsection{Incoherent Random Lasing; Theory}

  The first theoretical analysis of  random lasing 
dates back to 1967\cite{letokhov}. 
In the seminal paper of Letokhov \onlinecite{letokhov}
a random laser was modeled by an homogeneously pumped 
medium with scatterers that is contained  within a certain volume. 
Due to the excitation pump,  light diffusion inside the medium 
is accompanied by optical amplification. Lasing threshold is
then  determined 
from the condition that, in course of diffusion through
the volume,  light is amplified by a factor of $\sim 2$.   
Such an  amplification compensates for the losses from the
surface. The key idea of the theory \onlinecite{letokhov} 
is that multiple scattering leads to  {\em incoherent feedback}
by forcing the light to spend a relatively long time
inside the amplification region.  
Since light amplification  can be viewed as 
multiplication of the number of photons, the condition for
lasing is analogous to the criticality condition for  
chain reaction. 

Interest revival in diffusive propagation
of light in disordered gain media\cite{Genack94} has triggered
further advancement of the incoherent random lasing theory.  
In the  original work\cite{letokhov} 
it was assumed that the optical active centers are located inside the
{\em same} particles that scatter light. The immediate consequence
of this assumption is that a critical gain is independent of
the scatterers concentration.
To model realistic systems, the common situation
that was considered in later papers was a homogeneous laser dye solution
with suspended passive scattering microparticles. 
In this case, the description
of random lasing requires solving the 
system of coupled equations for 
light diffusion with amplification, and also for population inversion.
Various versions of such a system were 
invoked\cite{John96,Wiersma96,Genack96,Genack97,Balachabdran97}   
to study the steady state\cite{John96,Balachabdran97} as well as 
the dynamical\cite{Wiersma96,Genack96,Genack97}
properties of laser action in the scattering gain media.

In Ref. \onlinecite{John96} a generic model of a dye with 
singlet and triplet electronic excitations was considered.    
Taking into account the possibility that the emitted light 
from the  singlet band can serve as a pump for the triplet 
band yielded a bichromatic emission above a certain 
excitation threshold
with a ratio of emission peaks depending on the pump intensity 
and dye concentration. 
In Refs. \onlinecite{Wiersma96,Genack96,Genack97}
the temporal response of a disordered 
laser material to the incident pump 
(or pump and probe\cite{Wiersma96}) pulses was studied using
a Monte Carlo simulation of random walk of the pump and emitted
photons, which are interrelated through the population rate
equation. These studies revealed a rich emission dynamics
at excitation intensity close to  the threshold. Finally, in 
Ref. \cite{Balachabdran97} the Monte Carlo simulation was
employed to study  the emission spectral narrowing above the 
excitation threshold, as well as the input-output dependencies.

Summarizing, the original model of Ref. \onlinecite{letokhov} yields
a transparent and very appealing {\em quantitative} 
description  of the incoherent random lasing process.
However  it leaves out two important physical
effects: namely interference of diffusively moving waves, which
is a precursor of  photon localization
\cite{john85,john,Garcia91,Genack91,wiersma1,Wiersma97,Chabanov00} 
and    
strong interference-induced fluctuations of the local light
intensity, {\em i.e.} the mesoscopic 
phenomena\cite{alt91,berkovits,shapiro}. The  fundamental role of 
the two effects was not appreciated  back in 1967.

\subsection{Incoherent Random Lasing; Experiment}

The experimental results reported in the literature
\cite{Law94,wiersma,Sha94,Zhang95,Balachandran95,Noginov95,Martolrell96,Balachandran96,Sha96,Olivera97,Prasad97,Beckering97,Wang98,Soest99,Zach99,Shukri2000,Mar86,Mar91,Gou93,Hide96,Denton97} essentially confirm the theoretical predictions. Except for studies on powder grains of laser crystal materials \cite{Mar86,Mar91,Gou93} and $\pi$-conjugated polymer films,
\cite{Hide96,Denton97},
the majority of experiments\cite{Law94,wiersma,Sha94,Zhang95,Balachandran95,Noginov95,Martolrell96,Balachandran96,Sha96,Olivera97,Prasad97,Beckering97,Wang98,Soest99,Zach99,Shukri2000}
have used dye solutions as amplifying media. Colloidal particles
suspended in a solution served as random scatterers. In a typical
experiment a short pulse is used for
excitation. For below threshold pump intensities the emitted
pulse is long. It shortens drastically to $\lesssim 50ps$
as the laser threshold is exceeded. The emission spectrum
narrows to $\lesssim 20nm$. Numerical solution of the
the radiative transfer equations\cite{letokhov,John96,Wiersma96,Genack96,Genack97,Balachabdran97} yield the understanding 
of the dependence of the threshold energy on the
concentration of scatterers, concentration of dye, cavity size, etc. 
However, the most convincing proof that laser action is due 
to ``diffusive feedback'' stems from the fact that below 
threshold the presence
of scatterers has practically no effect on the emission.

\subsection{Coherent Random Lasing; Theory}

During the last decade the interplay of interference-induced 
localization effects 
and amplification of light in random media was the subject
of intensive theoretical 
\cite{zyuzin94,zyuzin95,zyuzin95',Pradhan94,Freilikh97,beenakker,Paasschens96,Jiang99,Jiang'99} studies. The theoretical 
prediction\cite{zyuzin94} 
 that longer diffusive trajectories due to gain result in 
 sharpening of  coherent backscattering cone
has been confirmed experimentally \cite{Wiersma95}.

The next step was incorporation of  disorder and interference
into the conventional system of Maxwell equations coupled with
the rate equations for the electronic population of a four-level-system
involved in  lasing. This program was realized in Ref. \cite{Sou2000}
for a model system representing a sequence of one dimensional layers
of random thickness.
One dimensional nature of the model and large (four times) contrast  in 
dielectric functions between the neighboring layers allowed the authors
to trace the crossover between the weak disorder {\em i.e.} long localization length (compared to the system size) and strong disorder regimes.
The simulations of Ref. \cite{Sou2000} have confirmed 
the basic idea of random lasing that increasing 
the disorder results in decreasing of the lasing threshold.
These simulations have also verified that lasing lines
in the emission spectrum have their origin in the localized
modes of the disordered one dimensional system.   
Further studies reported in  Ref. \cite{Sou2001} have indicated that
the distribution of electromagnetic field in the lasing mode is
practically the same as in the localized mode in the absence of
gain. The dependence of the emission spectrum on the gain frequency
was also studied in the simulations of Ref. \cite{Sou2001}. It
was demonstrated that, upon varying the gain frequency, 
the order in which different localized modes overcome the lasing 
threshold can change.

\subsection{Coherent Random Lasing; Experiment}

Recently a group of experiments was reported \cite{Cao99b,Cao2000a,Fro99,Fro99a}, which revealed  features of random lasing {\em qualitatively}
different from those reported previously in Refs. \cite{Law94,wiersma,Sha94,Zhang95,Balachandran95,Noginov95,Martolrell96,Balachandran96,Sha96,Olivera97,Prasad97,Beckering97,Wang98,Soest99,Zach99,Shukri2000}.


The disordered gain media in which
these measurements  were done were very different,  ranging from
$ZnO$ polycrystalline 
powders\cite{Cao99b,Cao2000a} to $\pi$-conjugated polymer films\cite{Fro99}, 
organic dyes-doped gel films\cite{Fro99} and dye-infiltrated opal
photonic crystals\cite{Fro99a}.
 However, a unifying feature of the results reported
using these media was the evolution of the emission spectra with
increasing excitation intensity.  Namely, the broad photoluminensce
band at low intensities first drastically narrows; as the excitation
intensity  was
increased even further, the emission spectrum transforms into a fine structure
that consists of a number of sharp ($\delta \lambda <$ 1nm) laser-like
emission lines.  
Using the photon counting statistics\cite{Zach2000},
the coherent nature of the
emitted light has been proved  for powders\cite{Caoprive}
as well as for polymer films\cite{Advanced}.


\begin{figure}
\narrowtext
{\epsfxsize=8.5cm
\centerline{\epsfbox{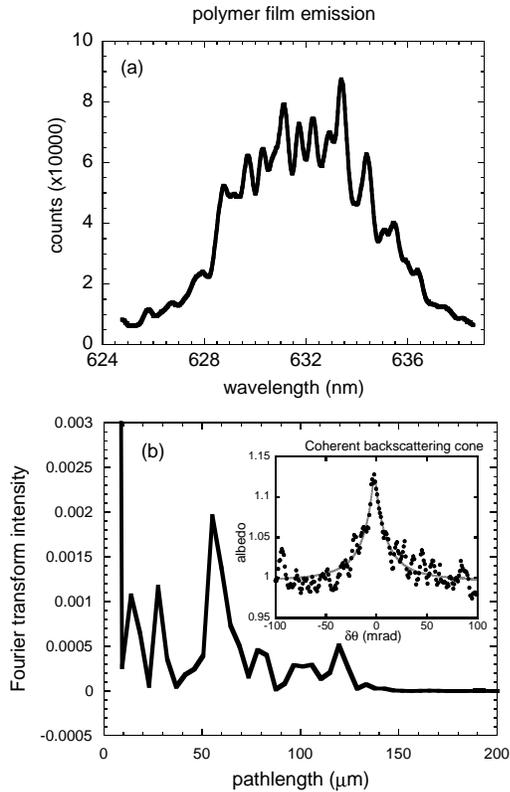}}}
\protect\vspace{0.5cm}
\caption{ (a) The measured emission spectrum of a DOO-PPV film 
above the threshold intensity for lasing. (b) Power Fourier transform of the emission spectrum in (a) . The inset shows the coherent backscattering cone from a bulk sample of DOOPPV.  }
\label{F:film}
\end{figure}
\noindent

The most comprehensive experimental study of random lasing thus
far has been performed by Cao {\em et. al.}
\cite{Cao99b,Cao2000a} on $Zn0$ powders  
that are extremely efficient scattering media\cite{Wiersma2000};
the light mean free path, $l^{\ast}$ extracted from coherent 
backscattering measurements
was $l^{\ast}= 0.8 \lambda$ in \cite{Cao99b} and 
$l^{\ast} = 0.5 \lambda$ in \cite{Cao2000a}.
On the basis of measurement of spatial distribution of the emitted
light
complemented with numerical 
simulations\cite{Cao2000a} it was demonstrated 
that the underlying mechanism of
lasing was formation of closed loop paths of light with a
characteristic size comparable to $\lambda$. These paths are
formed due to multiple scattering. It was argued 
in Ref. \cite{Cao99b} that, due to interference or, in other words,
localization  effects, 
such closed loop paths   
serve as resonant cavities for light. 
Thus, the formation of these tiny  cavities provides the  feedback
necessary for lasing. Correspondingly, the positions of the 
emission lines are determined by  particular configurations of
scatterers constituting each cavity~\cite{Cao2000a}.  

The evolution of the emission spectrum with pump power observed
in Ref. \cite{Cao99b} is in qualitative agreement with numerical
results of Ref. \cite{Sou2000} obtained for the domain of
small localization lengths.

\section{Experimental Results}

\subsection{Preliminary Discussion}

In this paper we demonstrate that despite the similarity in the
evolution of the emission spectra, the mechanism of random lasing in
$\pi$-conjugated polymer films is fundamentally different from that
reported for powders in Refs.\cite{Cao99b,Cao2000a}.  The difference
can be briefly formulated as follows: In the case of
Refs.\cite{Cao99b,Cao2000a} it was reported that {\em one} emission
line per random cavity was detected, whereas in polymer films one
resonator gives rise to a {\em number} of correlated emission lines.
This is because in our experiments with polymer films the relevant
random resonators are much bigger in size; this is not surprising
since the mean free path $l^{\ast}$ in the polymer film was measured
to be $l^{\ast}\sim 10 \lambda$.  Our principal finding is that all
random resonators are approximately {\em identical}.  We come to this
conclusion on the basis of the {\em regularity} in the experimentally
observed emission spectra.  Similarly to the case of electron
transport in which the underlying period has been experimentally
discerned from magnetoresitance {\em fluctuations}\cite{Wash91}, the
regularity in the {\em random} emission spectra is revealed here upon
Fourier analysis of the data.

\subsection{Random Spectra and Fourier Transforms}

Our strategy was as follows.  Fig.~\ref{F:film}(a) is a typical
emission spectrum above the laser threshold which was measured from a film excited with a stripe
illumination   of
$2mm \times 100\mu m$ that is  formed using a cylindrical lens. 
We note that in contrast,
the excitation spot in Refs. \cite{Cao99b,Cao2000a} was much smaller in
size $\sim 20\mu m$. Note also, that, despite the fact that 
in our case the two sizes of the excitation stripes differ by a 
factor of twenty. Nevertheless 
they  both are  {\em much bigger} than  $l^{\ast}$. 
Clearly therefore,
 the excited area cannot be considered as one dimensional 
\cite{Sou2000,Sou2001}.
The spectrum was measured at an excitation
intensity of $0.2\mu j$/pulse/$mm$ using $100 ps$ pulses from the second
harmonic of a $Nd:YAG$ regenerative amplifier that operated at $100 Hz$
repetition rate.  The specific polymer was poly(dioctyloxy) phenylene
vinylene, $DOOPPV$, which was synthesized by modifying a published
procedure\cite{BA95}.

\begin{figure}
\narrowtext
{\epsfxsize=8.5cm
\centerline{\epsfbox{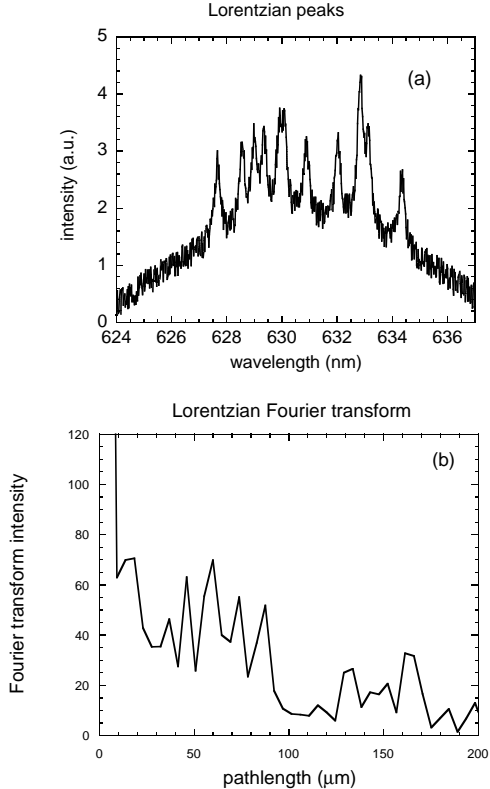}}}
\protect\vspace{0.5cm}
\caption{(a) A generated emission spectrum that consists of 12 Lorentzians.
(b)  Power Fourier transform of (a).}
\label{F:lorentzians}
\end{figure}

Fig.~\ref{F:film}(b) shows  the PFT
of Fig.~\ref{F:film}(a) 
and contains many apparent features.  In general, the presence of such
features in the PFT does not prove  regularity in the emission 
spectrum. To illustrate this,  we have generated a spectrum
very similar to Fig.~\ref{F:film}(a) 
by adding $12$ Lorentzians centered at
random wave lengths with  random linewidths.  
Fig.~\ref{F:lorentzians}(a) is a
numerically simulated spectrum and Fig.~\ref{F:lorentzians}(b) 
is the corresponding
PFT, which also shows many  apparent features.
Next we generated $125$ random spectra similar 
to Fig.~\ref{F:lorentzians}(a), and  averaged
their individual PFTs.  As would be expected, the
sharp PFT features vanished upon averaging.  
However when a similar procedure
was performed on $125$ {\em real} laser emission spectra obtained by 
shifting the
excitation stripe  on the surface of the polymer film, some features 
{\em persisted} in PFT;
moreover, they became much more pronounced, as seen in
Fig.~\ref{F:compare125FT} solid line. This supports the above
statement about universality made in the Introduction.

Similarly to the features in magnetoresitance, which serve as
fingerprints of actual disorder realization\cite{Wash91}, 
the peaks in a single
PFT reveal the properties of an individual random
resonator.  In Fig.~\ref{F:film}(b) the peaks allow us to determine the
size of a lasing resonator; more precisely the length  $L$ of the light path 
is calculated via the relation
 $L=\pi d/n$, where $d$ is the peak position in the
PFT and $n$ is the polymer index of refraction.  
Using $n=1.8$ for the {\em DOO-PPV} polymer  we get from
Fig.~\ref{F:film}(b)  a round trip length  $L=56 \mu m$ 
for the  peak at $d=18\mu m$, and $L=216\mu m$ for the peak at 
$d=60\mu m$, respectively. The smaller cavity corresponds to a length
 $L=90\lambda$,
which greatly exceeds the length scale $\sim \lambda$ as observed in
Refs. \onlinecite{Cao99b,Cao2000a}. 
It is also important to note that the
length $L$ also largely exceeds  
the light mean free path, $l^{\ast}$, which was
determined in DOO-PPV sample to be $l^{\ast}=5.2 \mu m$
or $8.3 \lambda$  from the width of the coherent
backscattering cone\cite{Huang} 
as shown in Fig.~\ref{F:film}(b), inset.

\noindent

\begin{figure}
\narrowtext
{\epsfxsize=8.5cm
\centerline{\epsfbox{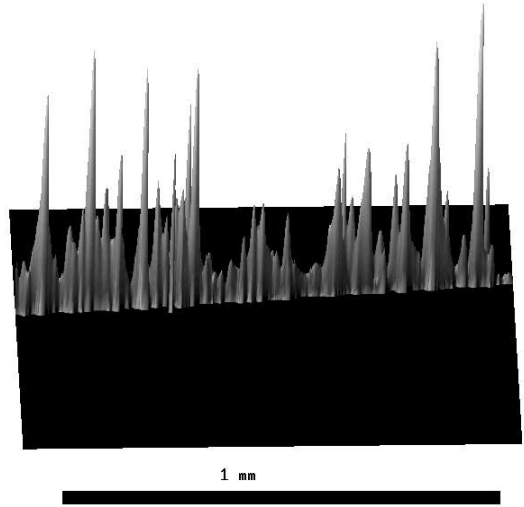}}}
\protect\vspace{0.5cm}
\caption{The measured spatial distribution of the emission
intensity in the DOO-PPV film under condition of random lasing.}  
\label{F:rubbersheet}
\end{figure}
\noindent

\subsection{Spatial Intensity Distribution}

Figure \ref{F:rubbersheet} is a magnified image of approximately half of
the excited area on the polymer film
 and measures the spatial distribution of the emission
intensity at an excitation intensity above the laser 
threshold.  The magnification is approximately $60$ times.
The emission intensity was measured through a long pass filter to 
remove the pump excitation
light and avoid saturating the image array.  The $z$-axis was computed
by  scaling the emission intensity of the red component of the 
color image.
A crude estimate of the distance between the bright
emitting  spots is
$40 \mu m$, which is roughly comparable to that
measured  in Ref.\cite{Cao2000a}.
However there is a stark difference between Fig. \ref{F:rubbersheet}
here and the corresponding spatial 
distribution in Ref. \onlinecite{Cao2000a},
namely that in our case there are {\em many more} bright spots than
spectral lines in the emission spectrum, as shown in 
Fig.~\ref{F:film}(a).  On the
other hand, bright spots {\em must} correspond to certain lines in the
emission spectrum.  Recall, that from the PFT a
characteristic size of the resonator $L=56 \mu m$ was inferred above. 
In the scale of Fig.~\ref{F:rubbersheet} this corresponds 
to a single bright
spot.  The total number of bright spots in Fig.~\ref{F:rubbersheet} 
is approximately three times the
number of emission lines in Fig. 2(a). Thus, 
the emission spectrum which contains a
limited number of lines reflects the emission from {\em many}
resonators.  The only resolution to this seemingly paradoxical
situation is that all the resonators that are responsible for lasing
in the polymer film
 are {\em approximately the same}.
This fact, and  the shape of the averaged PFT spectra
seen in Fig.~\ref{F:compare125FT} require a theoretical explanation,
which is given below.

\begin{figure}
\narrowtext
{\epsfxsize=8.5cm
\centerline{\epsfbox{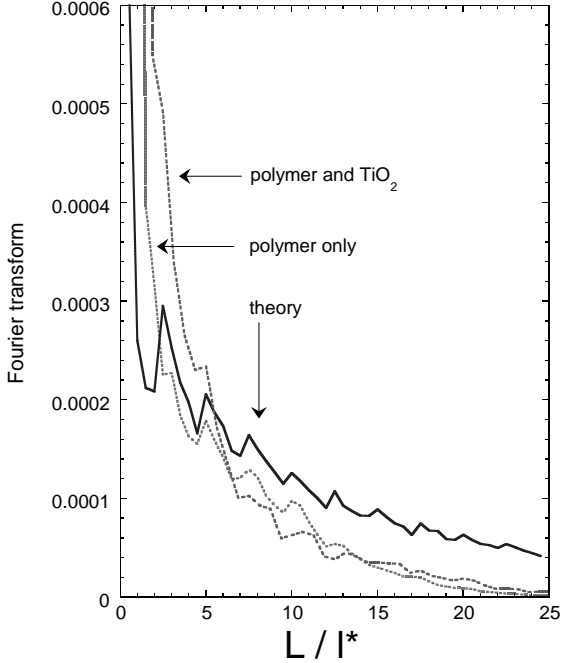}}}
\protect\vspace{0.5cm}
\caption{Average of 125 power Fourier transforms scaled by $l^{\ast}$.  The solid line is a theoretical model based on Eq. (\ref{E:avgFT})  superimposed on a smooth background; dotted lines are emission of pure polymer, and  polymer doped with $TiO_2$ scatterers, respectfully.}
\label{F:compare125FT}
\end{figure}
\noindent



\section{Theory}

We do not really know the actual shape of the random resonators
in the polymer films.
However, it is apparent that our case is very different
from the case in 
Refs.\cite{Cao99b,Cao2000a} since the round trip length, $L$
of a typical
resonator is much bigger than the mean free path $l^{\ast}$, which,
in turn, is much bigger than the wavelength $\lambda$.

Large values of $L$  suggest that the resonators
are formed due to the long-range inhomogeneities of the 
refraction index, $n(\mb{\rho})$. 
A plausible microscopic origin of these inhomogeneities  is the 
fluctuation $\delta h$ of the polymer 
film thickness, $h$, as illustrated in Fig.~1(a).
 Due to the waveguiding in the polymer film, the effective 
refraction index, 
$n_{\mbox{eff}}$, describing the in-plane propagation of light,
is related to $n$ as 
$n_{\mbox{eff}}=\sqrt{n^2-(p+1)^2\left(\frac{\lambda}{2h}\right)^2}$,
where $p$ is the number of the transverse waveguide mode. 
Therefore,
a local increase of the film thickness by $\delta h (\ll h)$ 
results in the
enhancement of $n_{\mbox{eff}}$ (see Fig. \ref{F:cartoon}(a))
by $\delta n_{\mbox{eff}}=(p+1)^2\lambda^2\delta h/4n h^3$.

In fact, long-range variations of the film thickness in {\em DOO-PPV}
polymer films were  reported in Ref. \cite{Frolov98}.
Since $\delta h \ll h$, and thus 
$\delta n_{\mbox{eff}} \ll n_{\mbox{eff}}$, the angle of total
internal reflection from the boundary (Fig. \ref{F:cartoon}(b)) is close to 
$90^{\circ}$. 

Thus, the microdisk-type
resonator that is formed due to  fluctuations of the film thickness (Fig. \ref{F:cartoon}(b)) can 
only support the ``whispering-gallery'', but not the ``bow-tie''
modes\cite{gmachl}.

\begin{figure}
\narrowtext
{\epsfxsize=8.5cm
\centerline{\epsfbox{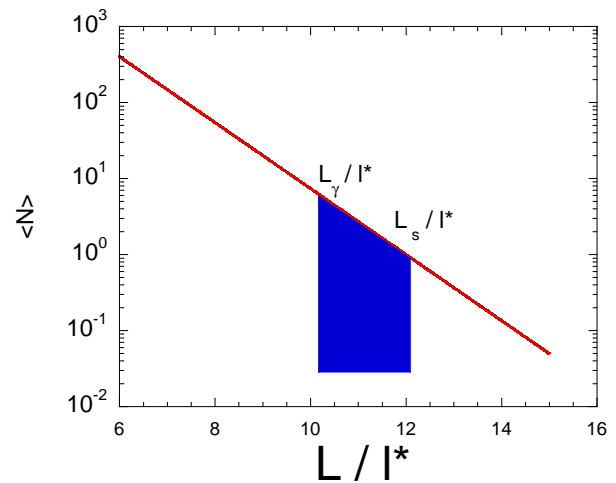}}}
\protect\vspace{0.5cm}
\caption{The average number of resonators $<N>$ that exist  within a sample with  area $S$ plotted versus the dimensionless pathlength $L / l^{\ast}$ (Eq. (\ref{average}).  $L_{\gamma}$ is the minimum length that is required  to overcome losses. The shaded area represents resonators that are able to lase.}
\label{F:expdecay}
\end{figure}

Below we demonstrate that the only assumption necessary to predict the
shape of the average PFT is that the formation of
resonators with  $L\gg l^{\ast}$ is highly
unlikely. In other words, we assume that the the areal {\em density},
$g(L)$, of resonators with a given length $L$ falls off exponentially 
with $L$. Quantitavely this fact can be expressed as follows
\begin{equation}  
\label{density}
g(L)=\frac{1}{l^{\ast 2}} \exp(-\varphi(L)),
\end{equation}
where the function $\varphi(L)$ increases with $L$, so that for $L\gg
l^{\ast}$ we have $\varphi(L)\gg 1$.  The actual form of the function
$\varphi(L)$ is determined by the type of  disorder in the film.
For a given area of the excitation spot, $S$, the maximal size,  $L_{S}$
of the resonator {\em present} within the area is determined by the
condition $Sg(L_{S})\sim 1$\cite{Raikh91}.  Since $g(L)$ 
is exponentially steep, this
 condition defines $L_{S}$ with  high accuracy. For values of
$L$ close to $L_{S}$, i.e. $\vert L-L_{S}\vert \ll L_{S}$, we can expand
$\varphi(L)$ as follows
\begin{equation}
\label{expand}
\varphi(L)=\varphi(L_{S})+\frac{L-L_{S}}{l_{0}},
\end{equation}
where $l_{0}^{-1}=\left(\partial\varphi /\partial L\right)_{L=L_{S}}$.
Using Eq. (\ref{expand}) the average number $N$ of resonators with the
path length $L$ within the illuminated area can be presented as
\begin{equation}
\label{average}
<N(L)>=\exp\left(\frac{L_{S}-L}{l_{0}}\right).
\end{equation}
The function $<N(L)>$ is shown in 
Fig.~\ref{F:expdecay}. At this point we note that for a given
gain, $\gamma$, there is a {\em minimum} resonator length,
$L_{\gamma}$, which is required to ensure lasing.  For example, in the
realization of a random resonator shown in Fig. 1
this length is determined by the condition
$\exp(\gamma L) F(L/L_{\mbox{c}})=1$, where the function $F$ accounts 
for losses
originating from the ``tunneling escape'' 
of light due to the curvature of the resonator 
perimeter\cite{Narimanov2000}. The characteristic length $L_c$ is
related to the variation $\delta n_{\mbox{eff}}$ as
$L_{\mbox{c}}=\lambda n_{\mbox{eff}}^{1/2}\left(\delta n_{\mbox{eff}}
\right)^{-3/2}\gg \lambda$. The asymptotics of the function 
$F$ at large values
of the argument is $F(z)=\exp(-2z/3)$.
In principle,   there exists another mechanism of losses originating
from generically non-circular shape of the 
resonator\cite{Narimanov2000}. However, for $\delta n_{\mbox{eff}}
\ll  n_{\mbox{eff}}$ the ``tunneling'' mechanism is dominant. 
The key point of our consideration is that, since the function $F$ 
is steep,
the condition $\exp(\gamma L_{\gamma})F(L_{\gamma}/L_c)=1$ defines 
the length,
$L_{\gamma}$ with {\em high accuracy}. 

For the illuminated stripe area
$S$,  lasing threshold is exceeded when the gain is sufficiently
high.
Quantitatively, this condition can be expressed as $L_{\gamma}<L_{S}$.
When  this condition is met, then  the lasing resonators 
 have lengths between $L_{\gamma}$ and $L_{S}$ (shaded area in Fig.~\ref{F:expdecay}).

The above arguments allow us to express the shape of the {\em average}
PFT. The PFT for a single resonator with 
a given length, $L$, can be presented as
\begin{equation} \label{E:expectedFT}
I(d)= \sum_{m} A_{m} \delta\left(mL-\frac{\pi d}{n} \right),
\end{equation}
where each $m$ is an harmonics index that represents a round trip in
the resonator. 
For all resonators $A_{m}$ fall off with
$m$, but  as gain starts to better compensate for
losses, the harmonics  amplitudes 
fall off more  slowly \cite{HT97,HT98}. 
The better gain compensates for losses, the slower is the decay
of $A_m$ with $m$. It is
usually approximated as $r^{m}$ \cite{HT98}, where $r$ is a factor
close unity; $(1-r)$ characterizes the proximity to the 
lasing threshold.   
Below laser threshold
Eq.~(\ref{E:expectedFT}) is exactly valid; 
however above laser threshold
the expected behavior of $A_{m}$ may change.  We have studied previously
\cite{RP2000} the change in the behavior of a single cavity 
crossing the lasing
threshold.  The results were that upon passing laser thresholdthe spacing between the components 
of the PFT stays the same, whereas
the factors
$A_{m}$ decrease in amplitude. In
Eq.~(\ref{E:expectedFT}) we used the fact that the spectrum of the
resonator is close to a delta function in the vicinity of the
threshold\cite{HT98}.
We note that the analysis of the coefficients
$A_{m}$ was recently suggested as a method to quantify resonator
losses below lasing threshold \cite{HT98}. Using
Eq.~(\ref{E:expectedFT}) the average (over excitation spots with area
$S$) PFT  takes the form
\begin{eqnarray} \label{E:avgFT}
<I(d)> \propto \sum_{m} A_{m} \int_{L_{\gamma}}^{\infty} dL \ \delta\left(mL-\frac{\pi d}{n} \right) <N(L)>  \nonumber \\
 \propto  \sum_{m} \frac{A_{m}}{m} \exp\left[  -m^{-1} \left(\frac{\pi d}{n l_{0}}\right ) \right] \Theta\left(\frac{ \pi d}{n} - mL_{\gamma} \right),
\end{eqnarray}
where $\Theta(x)$ is the step-function.
Eq. (\ref {E:avgFT}) indicates that the  average
PFT contains structures that are 
dictated by the length $L_{\gamma}$.  Therefore,
the presence of structures in Fig.~\ref{F:compare125FT} has a simple
physical interpretation.  Namely, that only  resonators with $L$ above a
certain, {\em well defined value} contribute to the laser emission spectrum. 
It is also seen  from Eq.~(\ref {E:avgFT})  that the dependence 
of $<I(d)>$ scales with $l_{0}$.

\subsection{Test of Universality}

The phenomenological parameter $l_{0}$ defined by Eq. (\ref{expand})
is the characteristics of particular type of the disorder in the 
film. Within a simple model that is illustrated in 
Fig. 1 with long-range fluctuations of the film thickness playing
the role of ``mirrors,''
this parameter should coincide
with the light mean free path, $l^{\ast}$,
which is determined by the short-range disorder. Indeed, as follows from
Eq.~(\ref{average}), upon increasing the path length by $\delta L$,
the number of resonators diminishes by a factor
\begin{equation}
\label{parameter}
\frac{<N(L+\delta L)>}{<N(L)>}=\exp\left(-\frac{\delta L}{l_{0}}\right).
\end{equation}
On the other hand, this reduction results from the fact that the
light is scattered away from the mirrors within

 the segment 
$\delta L$. The probability to {\em survive} this scattering is equal
to $\exp(-\delta L/\l^{\ast})$. Comparing this probability
with the r.h.s. of Eq. (\ref{parameter}), we conclude that 
$l_{0}=l^{\ast}$.

In order to test the prediction that the averaged PFT  
Eq. (\ref{E:avgFT}) scales with $l^{\ast}$, we have intentionally 
introduced
$TiO_2$ particles into the solution prior to the spin-casting process.
From the coherent backscattering measurements on bulk samples,  
we have established
that an additional short-range disorder caused a reduction of 
$l^{\ast}$ from $l^{\ast}=5.2\mu m$ in the pure film 
to $l^{\ast}= 4.1 \mu m$ in the $TiO_2$ doped polymer film. In 
Fig. (\ref{F:compare125FT})
we plot the PFT of the random laser spectra
for two polymer films, namely undoped and doped with $TiO_2$ 
particles, versus 
$d/\l^{\ast}$. Our first observation is that, whereas the peaks
in the PFTs of individual spectra appear to be
uncorrelated, the averaging over $125$ spots yields a remarkable
periodicity in the peak positions. This is in accordance with
the theoretical dependence that is implied in Eq. (\ref{E:avgFT}).
The first harmonic near $d/l^{\ast}=3$ differs by $20$ percent
from the data;
this discrepancy is likely from the discreteness of the PFT.
However, the discrepancy for the next five harmonics
is maximum $6$ percent.  
We thus conclude that the scaling is compelling.

In the same figure we also show a theoretical dependence, $<I(d)>$,
plotted using Eq. (\ref {E:avgFT}) that is superimposed on a smooth
background obtained from the structureless component of the
experimental PFTs in Fig. (\ref{F:compare125FT}). The only fitting
parameter in the theoretical $<I(d)>$ is the dimensionless ratio
$d/l^{\ast}$, which was set to be $2.9$.  It is seen that the
theoretical $<I(d)>$ curve exhibits the same saw-tooth structure as
the experimental curves.

\noindent
The fact that the peaks in the theoretical curve
diminish much slower with $d$ than in the experimental PFTs
can be accounted for incomplete averaging, since the higher is
the peak index, the higher is the  statistical error.

\section{Conclusions}

Until  now, two different regimes of random lasing have been 
addressed in the literature, namely, incoherent and coherent random 
lasing, respectively. They   correspond to different parameters domains
  of scattering media with optical gain.
Incoherent random lasing  occurs when, in the absence 
of gain, the propagation
of light is diffusive, {\em i.e.} $l^{\ast}\gg \lambda$.
Coherent random  lasing should take place
below or near the threshold of the Anderson localization transition, 
when $l^{\ast} \lesssim \lambda$. In  this paper
we demonstrate, however,  that in the presence of long-range inhomogeneities
of the refraction index with characteristic scale $L\gg \lambda$ the
coherent random lasing is also possible even when 
$l^{\ast}\gg \lambda$, so that {\em on average} 
the propagation of light is diffusive.
This is because inhomogeneities may trap  light 
within the scale $\sim L$ in a ``classical'' fashion (due to 
total internal reflections). If $L\gg l^{\ast}$, which is
the case for the polymer films that we have studied here, the
areal density of resonators that are able to trap  light
decreases rapidly with $L$ [roughly as $\exp(-L/l^{\ast})$].
On the other hand, at the lasing threshold, resonators
with the largest $L$ lase first. These two opposite trends 
effectively ``fix'' the value of $L$. As a result of $L$
being approximately fixed, the positions of the emission 
lines are  correlated. This correlation manifests itself
in the shape of the average PFT of the emission
spectra, which 
exhibits pronounced regular peaks.

We are grateful to H. Cao for illuminating discussions.
This work was supported by NSF  Grant No. DMR 9732820,
DOE Grant No. 96-ER45490, and the Petroleum Research Fund
Grant No. ACS-PRF 34302-AC6.

\end{document}